\documentclass[3p,times]{elsarticle}

\usepackage{amsmath}

\usepackage{amssymb}
\usepackage[figuresright]{rotating}

\usepackage[font=footnotesize,labelfont=bf]{caption}
\usepackage[font=footnotesize,labelfont=bf]{subcaption}

\usepackage[english]{babel}
\def\ds{\displaystyle} 
\usepackage{empheq,xcolor}

\begin{document}
	
	\begin{frontmatter}

		\title{Comments on the Properties of Mittag-Leffler Function}

		\author[Enea]{G. Dattoli \corref{cor}}
		\ead{giuseppe.dattoli@enea.it}
		
		\author[IJFPan]{K. Gorska }
		\ead{k.gorska80@gmail.com}
		
		\author[IJFPan]{A. Horzela }
		\ead{Andrzej.Horzela@ifj.edu.pl}
		
		\author[Enea,Unict]{S. Licciardi }
		\ead{silvia.licciardi@dmi.unict.it}
		
		\author[Unict]{R.M. Pidatella}
		\ead{rosa@dmi.unict.it}
		
		\address[Enea]{ENEA - Frascati Research Center, Via Enrico Fermi 45, 00044, Frascati, Rome, Italy}
		\address[IJFPan]{H.Niewodniczanski Institute of Nuclear Physics, Polish Academy of Sciences ,
			ul.Eliasza - Radzikowskiego 152, 31342 Krakow, Poland}
		\cortext[cor]{Corresponding author}
		\address[Unict]{Dep. of Mathematics and Computer Science, University of Catania, Viale A. Doria 6, 95125, Catania, Italy}
		
		\begin{abstract}
		 The properties of Mittag-Leffler function is reviewed within the framework of an umbral formalism. We take advantage from the formal equivalence with the exponential function to define the relevant semigroup properties. We analyse the relevant role in the solution of Schr\"{o}dinger type and heat-type fractional partial differential equations and explore the problem of operatorial ordering finding appropriate rules when non-commuting operators are involved.
         We discuss the coherent states associated with the fractional Sch\"{o}dinger equation, analyze the relevant Poisson type probability amplitude and compare  with analogous results already obtained in the literature. 
        \end{abstract}

		
	\end{frontmatter}
	
	\section{Introduction}

 In a previous paper \cite{Licciardi}, following a suggestion by Cholewinsky and Reneke \cite{Cholewinski}, it has been shown that, by a proper redefinition of the Newton binomial composition rule, the semigroup properties of exponential like functions can be recovered, along with the relevant consequences.\\

\noindent In particular, by introducing the Laguerre Binomial $ (l\!-\!b) $

\begin{equation} \label{eq1}
(x\oplus _{l} y)^{n} :=\sum _{r=0}^{n}\left(\!\!\begin{array}{c} {n} \\ {r} \end{array}\!\!\right)^{2}  x^{n-r} y^{r}                  
\end{equation} 
it has been proved that the function

\begin{equation}
{}_{l}e(x)=\sum_{r=0}^{\infty}\dfrac{x^r}{(r!)^2}
\end{equation}
called Laguerre exponential $l\!-\!e$, satisfies the following property

\begin{equation} \label{eq2}
_{l} e(x\oplus _{l} y)=\,_{l}e(y)\, {}_{l} e(x)                       
\end{equation} 
The $l\!-\!e$ can be recognized to be linked to the modified Bessel function of first kind  $I_{0} (x)$ according to the identity  

\begin{equation} \label{eq3} 
 _{l} e(x)  =I_{0} (2 \sqrt{x} )
\end{equation} 
We have denoted it as $l\!-\!e$ since it is an eigenvalue of the Laguerre derivative operator \cite{D.Babusci}

\begin{equation} \label{eq4} 
_{l} \partial _{x} =\partial _{x} \;x\;\partial _{x} 
 \end{equation}
According to the identity in eq. \eqref{eq2} the semigroup property of the ordinary exponential function is extended to the $l\!-\!e$, provided that the argument composition rule be replaced by the definition in eq. \eqref{eq1}. The use of the $l\!-\!b$ can exploited to define the Laguerre exponential through the following procedure of limit  \cite{Licciardi}. We note that the following ``generalizations'' of the properties of ordinary exponential

\begin{equation} \begin{split}\label{eq5} 
& \lim _{n\to \infty } \left(1\oplus_{l} \left(\frac{x}{n^{2} } \right)\right)^{n} ={}_{l} e(x)\\ 
& {}_{l} e:={}_{l} e(1)=\lim _{n\to \infty } \left(1\oplus_{l} \left(\frac{1}{n^{2} } \right)\right)^{n} =2.279585302336067...
\end{split}\end{equation} 
in which ${}_{l}e={}_{l}e(1)$ represents the $l$-Napier number and the limit can be viewed as the large index limit of ordinary Laguerre polynomials \cite{Licciardi,Andrews}. Within such a context, the $0$-order cylindrical Bessel function $J_0 (x)$
is given by 

\begin{equation} \label{eq6} 
\lim _{n\to \infty } \left(1\oplus_{l} \left(-\left(\frac{x}{2n} \right)^{2} \right)\right)^{n} =J_{0} (x) 
\end{equation} 
 The previous  remarks show that the point of view originated in ref. \cite{Cholewinski} and pursued in \cite{Licciardi} can be usefully extended to the theory of Mittag-Leffler ($M\!-\!L$) functions \cite{Mittag} and we will see the wealth of consequences of such a restyling.

\section{\textbf{Classical (one-parameter) Mittag-Leffler Function}}

The $M\!-\!L$ function has recently become of central importance in the theory of fractional derivatives \cite{Gorenflo}, this paper is devoted to a systematic investigation of the relevant properties, using different methods, including
operational, umbral and integral representation techniques.\\

\noindent $M\!-\!L$ is specified by the series expansion

\begin{equation} \label{eq7} 
E_{\alpha,1 } (x)=\ds \sum _{r=0}^{\infty }\dfrac{x^{r} }{\Gamma (\alpha \, r+1)} , \quad \forall x,\alpha \in \mathbb{R}, \alpha>0 
\end{equation} 
Its interest stems from the fact that it realizes to different special functions for different values of $\alpha$, e.g. for  $\alpha =1$ it yields the exponential, for $\alpha =\frac{1}{2}$ a combination of Dawson and Gaussian functions and for $\alpha =m, \; m\in \mathbb{Z}^+$ provides pseudo exponential functions and so on.\\
Unlike the ordinary exponential, to which it reduces for $\alpha =1$, the function $E_{\alpha,1 } (x)$ is such that

\begin{equation} \label{eq8} 
E_{\alpha,1 } (x+y)\ne E_{\alpha ,1} (x)\, E_{\alpha,1 } (y) 
\end{equation} 
according to the prescription of ref. \cite{Licciardi}, the redefinition of the Newton binomial  as 

\begin{equation} \begin{split}\label{eq9} 
& (x\oplus _{ml_{\alpha}} y)^{n} :=\sum _{r=0}^{n}\binom{n}{r} _{\alpha }\! x^{n-r} y^{r} , \\[1.2ex]
& \binom{n}{r} _{\alpha } :=\ds\dfrac{\Gamma (\alpha \, n+1)}{\Gamma \left(\alpha \, (n-r)+1\right)\Gamma (\alpha r+1)} 
 \end{split} \end{equation} 
allows the conclusion that

\begin{equation} \label{eq10} 
E_{\alpha,1 } (x\oplus _{ml_{\alpha}} y)=E_{\alpha,1 } (x)\, E_{\alpha,1 } (y)                \end{equation} 
The associated sin and cos-like functions defined by

\begin{equation}\begin{split} \label{eq11} 
& C_{\alpha,1 } (x)=\ds\frac{E_{\alpha,1 } (ix)+E_{\alpha,1 } (-ix)}{2} ,\\ 
& S_{\alpha,1 } (x)=\ds\frac{E_{\alpha,1 } (ix)-E_{\alpha,1 } (-ix)}{2\, i} 
 \end{split} \end{equation} 
also implying that

\begin{equation}
E_{\alpha,1 }(ix)=C_{\alpha,1 } (x)+i\;S_{\alpha,1 } (x)
\end{equation} 
 are characterized by the addition formulae

\begin{equation}\begin{split} \label{eq12} 
& C_{\alpha,1 } (x\oplus _{ml_{\alpha}} y)=C_{\alpha,1 } (x)\, C_{\alpha,1 } (y)-S_{\alpha,1 } (x)\, S_{\alpha,1 } (y), \\[1.2ex]
& S_{\alpha,1 } (x\oplus _{ml_{\alpha}} y)=S_{\alpha,1 } (x)\, C_{\alpha,1 } (y)+C_{\alpha ,1} (x)\, S_{\alpha,1 } (y)
 \end{split}\end{equation} 
resembling those of their circular counterpart. It is furthermore worth noting that, if $\alpha=n \in \mathbb{N}$, the $M\!-\!L$ function satisfies the eigenvalue equation

\begin{equation} \label{eq13} 
n ^{n } \left(x^{\frac{n -1}{n } } \frac{d}{dx} \right)^{n }  E_{n,1 } (\lambda x)=\lambda E_{n,1 } (\lambda x) 
\end{equation} 
It is therefore evident that by introducing the $M\!-\!L$ derivative operator

\begin{equation} \label{eq14} 
{}_{ml} \hat{D}_{x} =n ^{n } \left(x^{1-\frac{1}{n } } \frac{d}{dx} \right)^{n  }  
\end{equation} 
we find

\begin{equation} \label{eq15} 
E_{n,1 } \left(y\; {}_{ml} \hat{D}_{x} \right)E_{n,1 } (x)=E_{n,1 } \left(x\oplus _{ml_{\alpha}} y\right) 
\end{equation} 
Accordingly the operator $E_{n,1 } \left(y\; {}_{ml} \hat{D}_{x} \right)$ is a shift operator in the sense that it provides a shift of the argument of the $M\!-\!L$ function according to the composition rule established in eq. \eqref{eq9}.\\

\noindent It is also easily inferred from eq. \eqref{eq13} that\footnote{The last term is due to the fact that the fractional derivative in the sense of the Riemann-Liouville acts on one as follows: $\left(\dfrac{d}{dx}\right)^{\alpha}1=\dfrac{x^{-\alpha}}{\Gamma(1-\alpha)}.  $} 

\begin{equation} \label{eq16} 
\left(\frac{d}{dx} \right)^{\alpha }\! E_{\alpha,1 } (\lambda \, x^{\alpha } )=\lambda \, E_{\alpha,1 } (\lambda x^{\alpha } ) +\dfrac{x^{-\alpha}}{\Gamma(1-\alpha)}, \qquad \forall \alpha \in \mathbb{R}
\end{equation} 
%
\newline

 For future convenience  we introduce the umbral operator $\hat{c}$ \cite{Borel}, such that

\begin{equation} \label{eq17} 
\left. \hat{c}^{\,\mu }\dfrac{1}{\Gamma(z+1)}\right| _{z=0} :=\frac{1}{\Gamma (\mu +1)}  
\end{equation} 
in which $\varphi_{0}=\left. \dfrac{1}{\Gamma(z+1)}\right| _{z=0}$ is defined as "vacuum".  This term, borrowed from Physical language, is used to stress that the action of the operator $\hat{c}$, raised to some power,
is that of acting on an appropriate set of functions (in this case the Euler
Gamma function), by "filling" the initial "state" $\varphi_0 =\dfrac{1}{\Gamma(1)}$.\\
A fairly straightforward realization of the operator umbral operator is 

\begin{equation}\begin{split}\label{nota}
& \hat{c}^{\mu}\rightarrow e^{\mu \partial_z},\\
 & \varphi_z \rightarrow \dfrac{1}{\Gamma(z+1)}
\end{split}\end{equation}
so that

\begin{equation}\label{notab}
	\hat{c}^{\mu}\varphi_0 =e^{\mu \partial_z}\varphi_z \mid_{z=0}=\dfrac{1}{\Gamma(z+\mu+1)}\mid_{z=0}=\dfrac{1}{\Gamma(\mu+1)}
\end{equation}
Within the present context, the umbral realization of the $M\!-\!L$ function is\footnote{From this point on, for simplicity of writing, we omit the vacuum $\varphi_{0}$.} 

\begin{equation} \label{eq18} 
E_{\alpha,1 } (x)=\sum _{r=0}^{\infty }\left(\hat{c}^{\,\alpha } x\right)^{r} =\frac{1}{1-\hat{c}^{\,\alpha } x} 
\end{equation} 
The rigorous mathematical reasons underlying the operational calculus associated with the operator $\hat{c}$ trace back to the Borel transform technique and have been studied in detail in ref. \cite{Borel}. Noting furthermore that by standard Laplace transform we have

\begin{equation} \label{eq19} 
\frac{1}{1-\hat{c}^{\alpha } x} =\int _{0}^{\infty }e^{-s}  e^{\;\hat{c}^{\,\alpha } x\, s} ds                
\end{equation} 
 and that the expansion

\begin{equation} \label{eq20} 
e^{\;\hat{c}^{\alpha } x } =\sum _{r=0}^{\infty }\frac{\left(\hat{c}^{\alpha } x\right)^{r} }{r!}  =\sum _{r=0}^{\infty }\frac{x^{r} }{r!\Gamma (\alpha \, r+1)} 
\end{equation}
 yields a $0$-order Bessel-Wright function \cite{Wright}
 
\begin{equation} \label{eq21} 
W_{\alpha}^{(\mu)} (x)=\sum _{r=0}^{\infty }\frac{x^{r} }{r!\Gamma (\alpha \, r+\mu +1)}   
\end{equation} 
We can also conclude that the $M\!-\!L$ function is just the Borel transform of $W_{\alpha}^{(0)} (x)$ ,namely

 \begin{equation} \label{eq22}
E_{\alpha ,1} (x)=\int _{0}^{\infty }e^{-s}  W_{\alpha}^{(0)} (xs)ds                                            
\end{equation} 
A straightforward result, obtained for free from eqs. \eqref{eq18}-\eqref{eq19}, is

\begin{equation} \label{eq23} 
\left(\frac{d}{dx} \right)^{\!m} \!E_{\alpha,1 } (x)=\hat{c}^{\, \alpha\; m } \int _{0}^{\infty }e^{-s}  s^{m} e^{\;\hat{c}^{\,\alpha } x\, s} ds 
\end{equation} 
and noting that

\begin{equation} \label{eq24} 
\hat{c}^{\, \alpha\;m } e^{\hat{c}^{\,\alpha } \xi } =W_{\alpha}^{(\alpha \, m)} (\xi ) 
\end{equation} 
we can eventually end up with

\begin{equation} \label{eq25} 
\left(\frac{d}{dx} \right)^{m} E_{\alpha,1 } (x)=\int _{0}^{\infty }e^{-s}  s^{m} W_{\alpha}^{(\alpha \, m)}  (x\, s)ds             
\end{equation} 
The elements developed so far show that some of the properties of $M\!-\!L$ are obtained by the use of straightforaward means, which  provides the backbone of the forthcoming treatment.\\

We have shown that the umbral formalism is particularly useful for a straightforward handling of the $M\!-\!L$. The task of accomplishing the associated algebraic computation can be even more simplified by assuming for the $M\!-\!L$ an exponential umbral image, according to the prescription\footnote{It should be noted that the realization of umbral operator ${}_{\alpha}\hat{d}$ is the same as in eq. \ref{nota} (namely a shift differential operator),
the"vacuum" is however relized by a different function (namely by the ratio of two gamma functions. 
We have used a different notation, since to avoid a heavy notation, we will indicate the vacuum as $\varphi_0$}  


\begin{equation}\begin{split}\label{eq35}
& E_{\alpha,1}(x)=e^{\;{}_{\alpha}\hat{d} x},\\
&  {}_{\alpha}\hat{d} ^{\;\kappa}\left. \dfrac{\Gamma(z+1)}{\Gamma(\alpha z+1)}\right|_{z=0} =\dfrac{\Gamma(\kappa+1)}{\Gamma(\alpha \kappa+1)}
\end{split}\end{equation}
Such a restyling is as useful as the previous  and indeed we find

\begin{equation}\begin{split}\label{term}
\int_{-\infty}^{\infty}E_{\alpha,1}(-x^2)dx&=\sqrt{\pi}\; {}_{\alpha}\hat{d} ^{\;-\frac{1}{2}}\varphi_0=\sqrt{\pi}\dfrac{\Gamma\left( \dfrac{1}{2}\right) }{\Gamma\left( 1-\dfrac{\alpha}{2}\right) }=\\
& =\dfrac{\pi}{\Gamma\left( 1-\dfrac{\alpha}{2}\right) }
\end{split}\end{equation}
 and

\begin{equation}\label{mlb}
\int_{0}^{\infty}E_{\alpha,1}\left( -x^{\;\gamma}\right)dx=\dfrac{1}{\gamma}\dfrac{\Gamma\left( \dfrac{1}{\gamma}\right) \Gamma\left(1-\dfrac{1}{\gamma} \right) }{\Gamma\left( 1-\dfrac{\alpha }{\gamma}\right) }=\dfrac{\pi}{\gamma\sin\left(\dfrac{\pi}{\gamma} \right) }\dfrac{1}{\Gamma\left(1-\dfrac{\alpha}{\gamma} \right) } \qquad \forall \alpha,\gamma\in\mathbb{R}^+,\;\gamma>1,\; \frac{\alpha}{\gamma}\notin \mathbb{Z}^+
\end{equation}
Eqs. \eqref{term}-\eqref{mlb} can also be obtained by exploiting the properties of the umbral operator in $\hat{c}$-form, although with a slightly more cumbersome effort, as shown below.\\

According to the umbral definition \eqref{eq35} we can straightforwardly derive the Newton binomial \eqref{eq9} by noting that 

\begin{equation}
E_{\alpha,1}(x)E_{\alpha,1}(y)=e^{\;{}_{\alpha}\hat{d}_1 x}e^{\;{}_{\alpha}\hat{d}_2 y}=e^{\;{}_{\alpha}\hat{d}_1 x+{}_{\alpha}\hat{d}_2 y}
\end{equation}
which holds because ${}_{\alpha}\hat{d}_1,{}_{\alpha}\hat{d}_2$ are commuting operators separatly once acting on the vacua $1,2$ and specified below

\begin{equation}
{}_{\alpha}\hat{d}_b ^{\;\kappa}\left. \dfrac{\Gamma(z_b+1)}{\Gamma(\alpha z_b+1)}\right|_{z=0} =\dfrac{\Gamma(\kappa+1)}{\Gamma(\alpha \kappa+1)}, \;\;\; b=1,2
\end{equation}
and executing the explicity computation we obtain

\begin{equation}
E_{\alpha,1}(x)E_{\alpha,1}(y)=
\sum_{n=0}^{\infty}\dfrac{1 }{n!}
\left( {}_{\alpha}\hat{d}_1 x+{}_{\alpha}\hat{d}_2 y\right)^n =E_{\alpha,1 } (x\oplus _{ml_{\alpha}} y)
\end{equation}

\subsection{\textbf{Generalized (two-parameters) Mittag-Leffler Function}}


 We will henceforth adopt for the $M\!-\!L$  the definition given in \cite{Mittag} and write 
 
\begin{equation} \label{eq26} 
E_{\alpha ,\, \beta } (x)=\sum _{r=0}^{\infty }\frac{x^{r} }{\Gamma (\alpha \, r+\beta )}  =\int _{0}^{\infty }e^{-s}  W_{\alpha}^{(\beta -1)} (x\, s)ds, \qquad \forall x \in \mathbb{R}, \forall \alpha,\beta\in\mathbb{R}^+ 
\end{equation} 
According to the umbral definition in terms of the $\hat{c}$-operator, we simply obtain 

\begin{equation} \label{eq27}
E_{\alpha ,\, \beta } (x)=\frac{\hat{c}^{\,\beta -1} }{1-\hat{c}^{\,\alpha } x}                          
\end{equation} 
or, using an exponential umbral image, we write

\begin{equation}\begin{split}\label{dab}
& E_{\alpha,\;\beta}(x)=e^{\;{}_{\alpha,\;\beta}\hat{d}\;x},\\
& {}_{\alpha,\;\beta}\hat{d}^{\;\kappa}=\dfrac{\Gamma(\kappa+1)}{\Gamma(\alpha\kappa+\beta)}
\end{split}\end{equation}
The extra term containing $\beta$ does not provide any substantive difference with respect to the previous discussion and, e.g., the relevant modified binomial coefficient is defined as

\begin{equation} \label{eq28}
\binom{n}{r}_{\alpha ,\;\beta } :=\dfrac{\Gamma (\alpha \, n+\beta )}{\Gamma \left(\alpha \, (n-r)+\beta \right)\Gamma (\alpha r+\beta )}        
\end{equation} 
and consequently

\begin{equation}
(x\oplus _{ml_{\alpha,\beta}} y)^{n} :=\sum _{r=0}^{n}\binom{n}{r} _{\alpha,\beta }\! x^{n-r} y^{r}
\end{equation}
To give an example of the reliability of the method we note, e.g., that integrals of the type 

\begin{equation} \label{eq29}
\int _{-\infty}^{\infty }E_{\alpha ,\, \beta } (-x^{2} )\; dx=c^{\;\beta -1} \int _{-\infty}^{\infty }\frac{dx}{1+\hat{c}^{\alpha } x^{2} }  =\frac{\pi }{2} \hat{c}^{\beta -\frac{\,\alpha }{2} -1} =\frac{\pi }{ \Gamma \left(\beta -\frac{\alpha }{2} \right)}     
\end{equation} 
are easily obtained by noting that they derive from the identity 

 \begin{equation}\label{integr}
  \int _{-\infty }^{\infty }\frac{dx}{1+a^{\alpha } x^{2} }  =\pi \; a^{-\frac{\alpha }{2} } 
 \end{equation}
   upon replacing $a$ with $\hat{c}$ and after applying the so far discussed rules.\\
   
   The same procedure extended to the rapresentation \eqref{dab} yields
   
   \begin{equation}
   \int_{-\infty }^{\infty }E_{\alpha,\beta}(-x^2)dx=\int_{-\infty }^{\infty }e^{-{}_{\alpha,\;\beta}\hat{d}\;x^2}dx=\sqrt{\pi}\left( {}_{\alpha,\;\beta}\hat{d}\right) ^{-\frac{1}{2}}=\frac{\pi }{ \Gamma \left(\beta -\frac{\alpha }{2} \right)} 
   \end{equation}

\section{The Mittag-Leffler Functions and Fractional Calculus}

 Let us now consider the following Cauchy problem \cite{FFP}
 
\begin{equation}\label{cauchy} 
\left\lbrace  \begin{array}{l} \partial _{t}^{\alpha } F(x,t)=\partial _{x}^2 F(x,t)+\dfrac{t^{-\alpha}}{\Gamma(1-\alpha)}f(x) \\[1.2ex]
 F(x,0)=f(x) \end{array}\right.
 \end{equation} 
defining a $PDE$ resembling a kind time-fractional diffusive equation. According to the previous discussion and to the fact that the $M\!-\!L$,  ''$E_{\alpha ,1} (t^{\alpha } ) $'', is an eigenfunction of the fractional derivative operator according to the definition \eqref{eq16},  if we trust the formalism developed so far, we can obtain the relevant solution in the form \cite{FFP}

\begin{equation} \label{eq31} 
F(x,t)=E_{\alpha ,1} (t^{\alpha } \partial _{x} ^2)\, f(x)
\end{equation} 
where $E_{\alpha ,1} (t^{\;\alpha } \partial _{x}^2 )\, $is the evolution operator for the problem under study.\\

\noindent The relevant action on the initial function can be espressed as 

\begin{equation}\label{Fxt}
F(x,t)=\dfrac{1}{\sqrt{2\, \pi } } \int _{-\infty }^{+\infty }E_{\alpha ,1}(-t^{\alpha } k^2 )\, \tilde{f}(k)\, e^{i\, k\, x} dk 
\end{equation}
where $\tilde{f}(k)$ is the Fourier transform of $f(x)$\footnote{We observe that eq. \eqref{Fxt} can be recast in terms of Levy distribution according to ref. \cite{FFP}.}.\\

\begin{figure}[htp]
	\centering
	\begin{subfigure}[c]{0.48\textwidth}
		\includegraphics[width=0.9\linewidth]{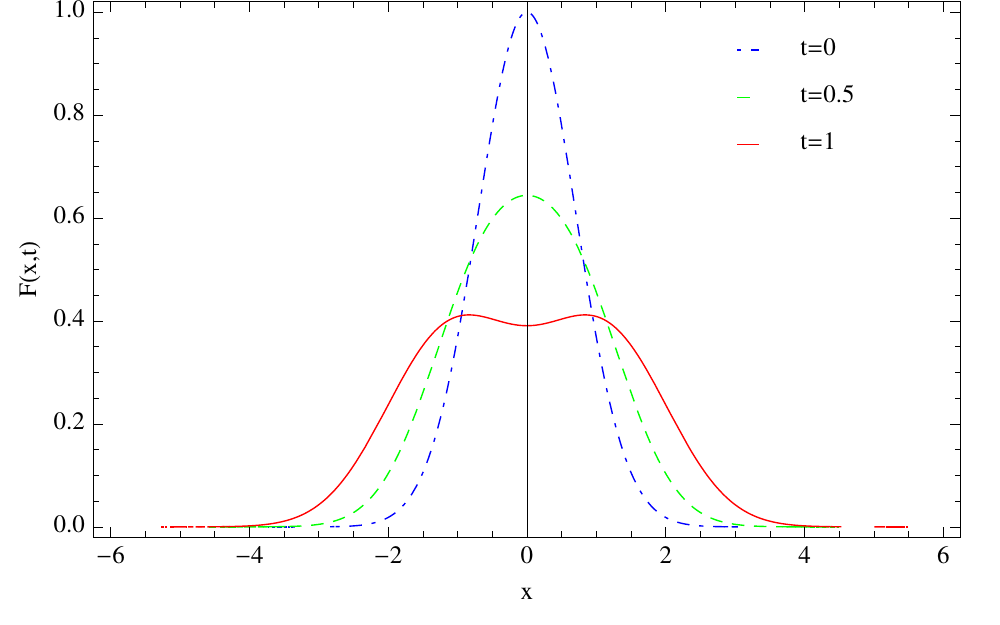}
		\caption{$\alpha=1.5$.}
		\label{Fig1a}
	\end{subfigure}
	\begin{subfigure}[c]{0.48\textwidth}
		\includegraphics[width=0.9\linewidth]{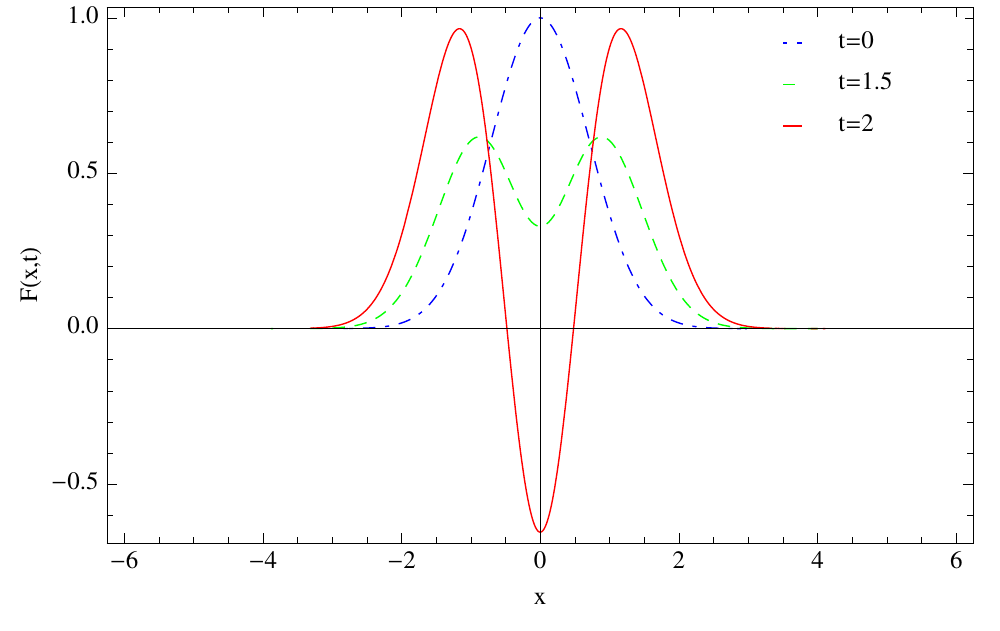}
		\caption{$\alpha=3.5$.}
		\label{Fig1b}
	\end{subfigure}
	\\[3mm]
	\caption{Solution of eq. \eqref{cauchy} for $f(x)=e^{-x^{2}}$.}\label{Fig1} 
\end{figure}

\noindent Examples of solutions are reported in Figs. \ref{Fig1} 

which clearly display a behaviour which is not simply diffusive but also anomalous\footnote{The variance of $x$ depend on time and is proportional to $\dfrac{2 t^{\;\alpha}}{\Gamma(1+\alpha)}$ according to eq. (20) in  \cite{FFP}.} (for further comments see ref. \cite{FFP}). \\

The introduction of the Pseudo-Exponential Operator ($PEO$), "$E_{\alpha ,1} \left( t^{\alpha } \partial _{x}^2 \right) $",  is of central importance for our forthcoming discussion,  its role and underlying computational rules should be therefore carefully understood.\\

We remind therefore that, regarding ordinary evolution problems, ruled by an equation of the type

\begin{equation} \label{eq33} 
\left\lbrace  \begin{array}{l} \partial _{t} F(x,t)=\hat{O}\, F(x,t) \\[1.2ex] F(x,0)=f(x) \end{array} \right. 
\end{equation} 
with $\hat{O}$ being  a not specified operator acting on the $x$ variable, the semigroup property of the exponential ensures that the associated evolution operator 

\begin{equation}
\hat{U}(t)=e^{\;t\, \hat{O}}
\end{equation}
 produces a shift of the time variables which can be expressed as

\begin{equation}\label{Oop}
 \hat{U}(t_{2} )F(x,t_{1} )=F(x,\, t_{2} +t_{1} )
\end{equation}
The extension of such a property to evolution driven by $PEO$ requires some caution. \\

\noindent We first note that if\footnote{The integral representation of $\hat{U}_{\alpha}=E_{\alpha,1}(t^{\;\alpha}\hat{O})$ is given in eq. (12) in \cite{FFP}.}


\begin{equation}\label{sS}
f(x)=S(x,\bar{t})=E_{\alpha,1}(\bar{t}^{\;\alpha}\hat{O})s(x)
\end{equation} 
with $s(x)$ initial condition,  then 

\begin{equation}\label{somml}
E_{\alpha,1}(t^{\;\alpha}\hat{O})f(x)=E_{\alpha,1}(t^{\alpha}\hat{O})\left( E_{\alpha,1}\left(\bar{t}^{\;\alpha}\hat{O}\right)s(x)  \right)=S(x,t\oplus_{ml_{\alpha}}\bar{t}) 
\end{equation}
The role of $PEO$ as time translation operator holds therefore in a broader sense.\\

%
%
%

Before proceeding further, let us consider the following operator

\begin{equation}\begin{split}
& \Phi_{\alpha}\left( \hat{A},\hat{B}\right)=E_{\alpha,1} \left( \hat{A}+\hat{B}\right),\\
& \left[  \hat{A},\hat{B}\right] =\hat{A}\hat{B}-\hat{B}\hat{A}=k \hat{1}
\end{split}\end{equation}
where the operators  $\hat{A}$ and $\hat{B}$ are not commuting each other, but their commutator, $k$, commutes either with $\hat{A}$ and $\hat{B}$ ($k,\hat{1}$ are  a non-necessarily real number and the unit operator, respectively).\\

\begin{figure}[htp]
	\centering
	\begin{subfigure}[c]{0.48\textwidth}
		\includegraphics[width=0.9\linewidth]{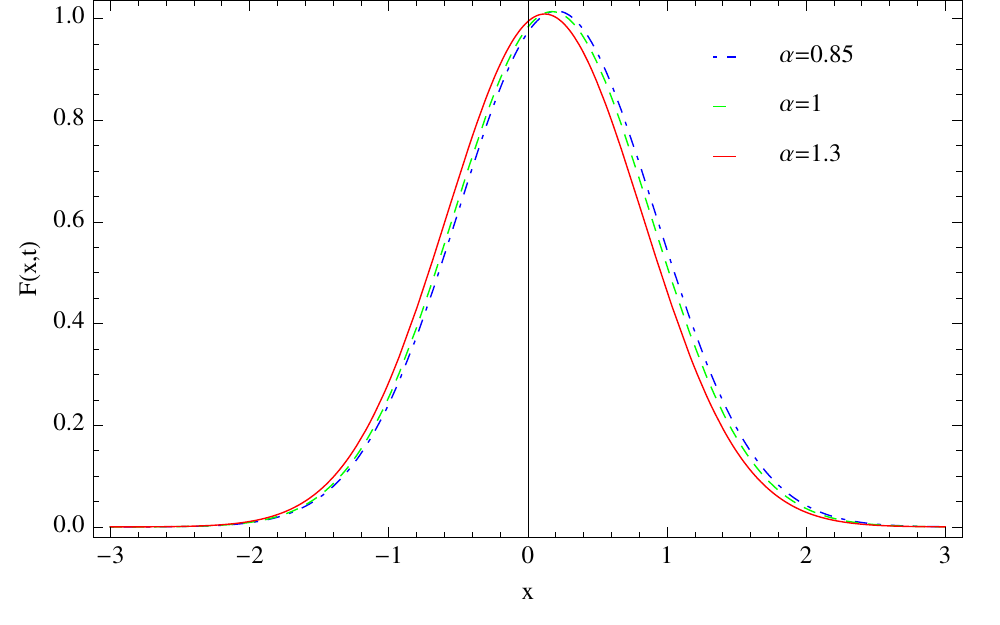}
		\caption{$t=0.5$.}
		\label{Fig2a}
	\end{subfigure}
	\begin{subfigure}[c]{0.48\textwidth}
		\includegraphics[width=0.9\linewidth]{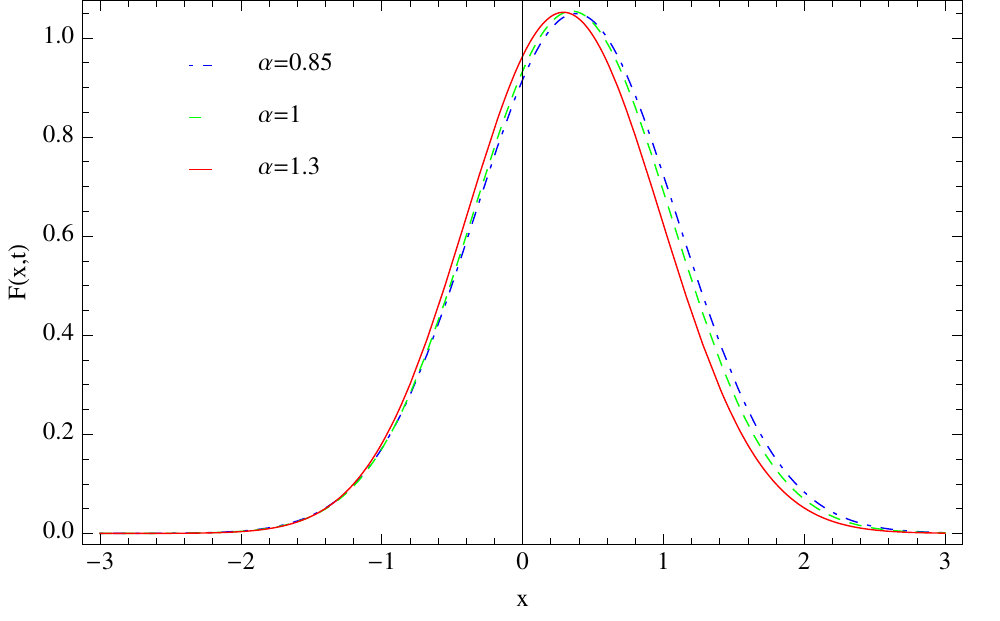}
		\caption{$t=1$.}
		\label{Fig2b}
	\end{subfigure}
	\begin{subfigure}[c]{0.48\textwidth}
		\includegraphics[width=0.9\linewidth]{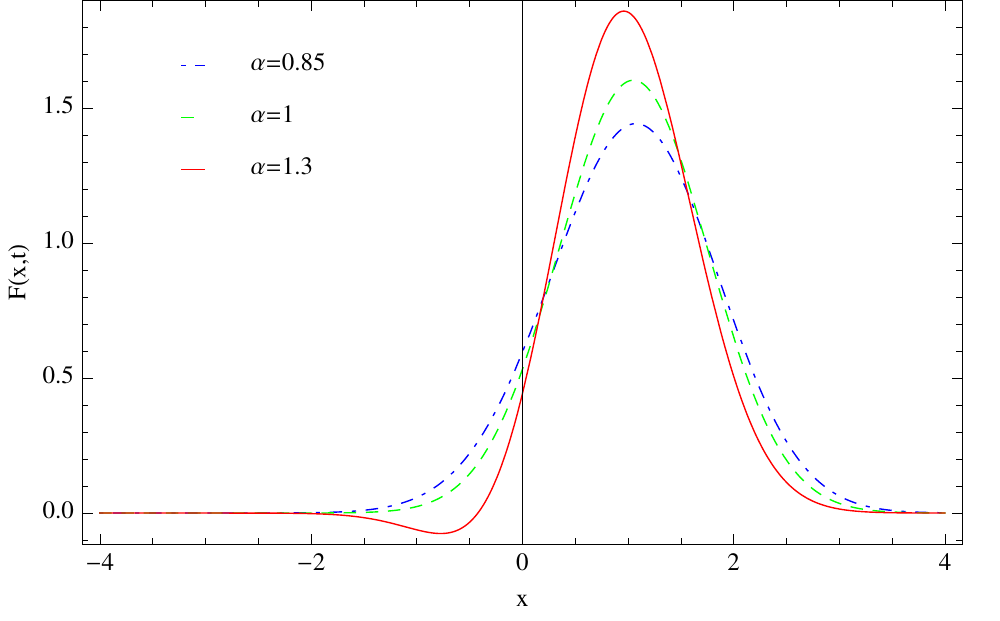}
		\caption{$t=3$.}
		\label{Fig2c}
	\end{subfigure}
	\\[3mm]
	\caption{Cauchy Fractional Problem Solution $F(x,t)$ vs $x$ for different $t$ and $\alpha$ values.}\label{Figure2} 
\end{figure}

\noindent In the case of ordinary exponential we find, by the use of the Weyl identity \cite{Dattoli}, the ordering rule

\begin{equation}\label{rulesComp}
e^{\hat{A}+\hat{B}}=e^{\hat{A}}e^{\hat{B}}e^{-\frac{k}{2}}
\end{equation}
In close analogy, using eq. \eqref{eq35} to write the $M\!-\!L$ function, we get

\begin{equation}\begin{split}
& E_{\alpha,1} \left( \hat{A}+\hat{B}\right)=e^{\; _{\alpha}\hat{d}  \left( \hat{A}+\hat{B}\right)}=e^{\;_{\alpha}\hat{A}+_{\alpha}\hat{B}},\\
& _{\alpha}\hat{S}=\;_{\alpha}\hat{d}
\hat{S}\end{split}\end{equation}
and noting that

\begin{equation}
\left[ _{\alpha}\hat{A},\; _{\alpha}\hat{B}\right] =\; _{\alpha}\hat{d} ^{\;2} k 
\end{equation}
we can therefore write

\begin{equation}
e^{\;_{\alpha}\hat{A}+_{\alpha}\hat{B}}=e^{\;_{\alpha}\hat{A}}e^{\;_{\alpha}\hat{B}}e^{-\frac{k}{2}\;_{\alpha}\hat{d} ^{\;2}}
\end{equation}
To understand the meaning  of the previous result we consider the following PDE

\begin{equation}\label{DiffEq2}
\left\lbrace \begin{array}{l} \partial_{t}^{\alpha}F(x,t)=\left(a\; x-b\;\partial_{x} \right) F(x,t)+\dfrac{t^{-\alpha}}{\Gamma(1-\alpha)} \\[1.2ex] F(x,0)=e^{-x^2}
 \end{array} \right. 
\end{equation}
Being the initial function just a constant, we can cast the relevant solution in the form

\begin{equation}\label{newsolD}
F(x,t)=e^{\;_{\alpha}\hat{d}\;t^{\alpha}\left(a x-b\partial_{x} \right)}e^{-x^2}
\end{equation}
by setting therefore

\begin{equation}\begin{split}
& _{\alpha}\hat{A} = {}_{\alpha}\hat{d}\;t^{\alpha}a\;x\\
& _{\alpha}\hat{B} =- {}_{\alpha}\hat{d}\;t^{\alpha}b\;\partial_x
\end{split}\end{equation}
being

\begin{equation}
\left[ _{\alpha}\hat{A},  {}_{\alpha}\hat{B}\right] =\left( _{\alpha}\hat{d}\right)^2 a\; b\; t^{2\alpha} 
\end{equation}
we end up with

\begin{equation}\label{ddd}
F(x,t)=e^{\;_{\alpha}\hat{d}\; t^{\alpha}\;a\,x - \frac{_{\alpha}\hat{d}^{\;2}}{2}\; t^{2\alpha}\;a\;b }e^{- _{\alpha}\hat{d}t^{\alpha}b\; \partial_x}e^{-x^2}=e^{\;a\;x\;  \left( _{\alpha}\hat{d} t^{\alpha}\right) -	\frac{ab}{2}\left( _{\alpha}\hat{d}\; t^{\alpha}\right)^2 }e^{-\left( x-\;{}_{\alpha}\hat{d}\;t^\alpha \;b\right)^2 }
\end{equation}
The r.h.s can be treated by using the properties of Hermite Kamp\'e de F\'eri\'et polynomials

\begin{equation}
H_n (x,y)=n!\sum_{r=0}^{\lfloor\frac{n}{2}\rfloor}\dfrac{x^{n-2r}y^r}{(n-2r)!r!}
\end{equation}
whose generating function writes

\begin{equation}
\sum_{n=0}^{\infty}\dfrac{t^n}{n!}H_n (x,y)=e^{xt+yt^2}
\end{equation}
accordingly we end up with the following explicit solution

\begin{equation}\begin{split}\label{solDF2}
F(x,t)&= e^{-x^2}
e^{x\;  (a+2b)\left( _{\alpha}\hat{d} t^{\alpha}\right) -	\left( \frac{a}{2b}+1\right) \left( _{\alpha}\hat{d}\; t^{\alpha}\right)^2 b^2}=
e^{-x^2}\sum_{r=0}^{\infty}\dfrac{\left( _{\alpha}\hat{d}t^{\alpha}\right)^r }{r!}H_r \left( x(a+2b),-b^2	\left( \frac{a}{2b}+1\right)\right)=\\
& = e^{-x^2}\sum_{r=0}^{\infty}\dfrac{t^{\alpha r}}{r!}\dfrac{\Gamma(r+1)}{\Gamma(\alpha r+1)}H_r \left( x(a+2b),-b^2	\left( \frac{a}{2b}+1\right)\right)=\\
& =e^{-x^2}\sum_{r=0}^{\infty}\dfrac{t^{\alpha r}}{\Gamma(\alpha r+1)}H_r \left( x(a+2b),-b^2	\left( \frac{a}{2b}+1\right)\right)
\end{split}\end{equation}
We point out that eq. \eqref{solDF2} can be also obtained from eq. \eqref{newsolD}, with the integral form of the  evolution operator \eqref{sS} given in eq. (12) in \cite{FFP}, or the $\hat{c}$-umbral representation of the eqs. \eqref{eq18}-\eqref{eq19} and the exponential generating function in \eqref{ddd}. The solution $F(x,t)$ of the Cauchy fractional problem \eqref{DiffEq2}, in eq. \eqref{solDF2}, has been plotted vs. $x$ for different times in Figs. \ref{Figure2}.\\ 

The relevant meaning and the link with fractional Poisson processes, as well, will be discussed in the forthcoming sections. 

\section{Non-Standard Statistical Effects and Mittag-Leffler Functions}

In this section we will consider a time fractional Schroedinger equation regarding a process implying the emission and absorption of photons. We will assume that the relevant dynamics is ruled by the $M\!-\!L$ Schroedinger equation

\begin{equation}\begin{split}\label{CrAnn}
& i^{\alpha}\;\partial_{t}^{\alpha}\mid \Psi \;\rangle= \hat{H}\mid \Psi \;\rangle+i^{\alpha}\dfrac{t^{-\alpha}}{\Gamma(1-\alpha)}\mid \Psi(0)\;\rangle, \\
& \hat{H}=i^{\alpha}\;\Omega\left( \hat{a}-\hat{a}^{+}\right)
\end{split}\end{equation}
where  $\hat{a},\hat{a}^{+}$ are annihilation, creation  operators satisfying the commutation bracket "$\left[ \hat{a},\hat{a}^{+}\right] = \hat{1}$", \cite{Louiselle}. The constant $\Omega$ in eq. \eqref{CrAnn} has the dimension of $t^{-\alpha}$.\\

\noindent It should be noted that in writing the fractional version of the Schroedinger equation, we have assumed that the l.h.s. operator is 
"$\left( i\dfrac{d}{dt}\right) ^{\alpha} $" and not "$i\left( \dfrac{d}{dt}\right) ^{\alpha} $". There is not a general consensous on such a choice \cite{Naber,N.Laskin}\footnote{Although the paper in ref. \cite{N.Laskin} deals with  a fractional spatial form of the Schroedinger equation, the points raised are relevant also for the topics treated here.}, which should be driven by an educated  guess on the concept of the "hermiticity" of the operators entering a fractional Schroedinger equation. The present choice retains the hermiticity in the usual sense of the left hand side, namely

\begin{equation}
 \left(\left(  i\dfrac{d}{dt}\right) ^{\alpha} \right) ^+=\left( i\dfrac{d}{dt}\right) ^{\alpha} 
\end{equation}
but regarding the right hand side

\begin{equation}
\hat{H}^+ \neq\hat{H}
\end{equation}
 Further comments on these points will be provided in the concluding section. \\
 The choice of the appropriate system of units in which the equation should be written is a further  aspect of the problem to be carefully considered, but will not be discussed here. For further comments we address the reader to ref. \cite{Naber}, where the rescaling of the variables in terms of Planck units has been proposed.\\

\noindent If we work in a Fock basis and choose the "physical" vacuum (namely the state of the quantized electromagnetic field with no photons) as the initial state of our process, namely 

\begin{equation}
\mid \Psi\; \rangle   \mid_{t=0}=\mid 0\;\rangle
\end{equation}
we can understand how the field ruled by a fractional Schroedinger equation evolves from the vacuum. The comparison with the ordinary Schroedinger counterpart is interesting because, in this case, the field is driven from the vacuum into a coherent state, displaying an emission process in which the photon counting statistics follows a Poisson distribution \cite{Louiselle}.\\

\noindent The solution of the evolution problem in eq. \eqref{CrAnn} can, according to the rules developed in the previous sections\footnote{In particular, in according to the eq. \eqref{rulesComp}, we assume that $\hat{A}=-\left( \; _{\alpha}\hat{d}\; t^{\alpha}\; \Omega\right)\;\hat{a}^{+}$ and $\hat{B}=\left( \; _{\alpha}\hat{d}\; t^{\alpha}\; \Omega\right)\;\hat{a}$. Furthermore, since $\left( \; _{\alpha}\hat{d}\; t^{\alpha}\; \Omega\right)\;$ and $\hat{a},\hat{a}^+$ are commuting operators, we can set $\left[ \hat{A},\hat{B}\right] =\left( \; _{\alpha}\hat{d}\; t^{\alpha}\; \Omega\right)^2 $.}, be written as

\begin{equation}\begin{split}\label{eq52}
& \mid \Psi \;\rangle=e^{\; _{\alpha}\hat{d}\;t^{\alpha}\; \Omega\;   \left( \hat{a}-\hat{a}^{+}\right)}\mid 0\;\rangle=\\
& = e^{- \frac{ \left( \; _{\alpha}\hat{d}\; t^{\alpha}\; \Omega\right) ^2}{2} }e^{-\left( \; _{\alpha}\hat{d}\; t^{\alpha}\; \Omega\right)\;\hat{a}^+}e^{\left( \; _{\alpha}\hat{d}\; t^{\alpha}\; \Omega\right)\; \hat{a}}\mid 0\;\rangle
\end{split}\end{equation}
The use of the identities 

\begin{equation}\begin{split}
& (\hat{a}^+)^n \mid 0 \;\rangle= \sqrt{n!}\mid n\;\rangle,\\
& \hat{a} \mid 0\;\rangle=0
\end{split}\end{equation}
and the remarks of the previous section allows the derivation of the explicit solution in eq. \eqref{eq52} as

\begin{equation}
\mid \Psi \;\rangle=e^{- \frac{ \left( \; _{\alpha}\hat{d}\; t^{\alpha}\; \Omega\right) ^2}{2} }e^{-\left( \; _{\alpha}\hat{d}\; t^{\alpha}\; \Omega\right)\hat{a}^+}\mid 0\;\rangle= e^{- \frac{ \left( \; _{\alpha}\hat{d}\; t^{\alpha}\; \Omega\right) ^2}{2} }\sum_{n=0}^{\infty}\dfrac{\left( -\; _{\alpha}\hat{d}\; t^{\alpha}\; \Omega\right)^n }{\sqrt{n!}}\mid n\;\rangle
\end{equation}
The probability amplitude of finding the state $\mid \Psi >$ in a photon number state $\mid m>$ is just given by

\begin{equation}
\langle m\mid \Psi\;\rangle= e^{- \frac{ \left( \; _{\alpha}\hat{d}\; t^{\alpha}\; \Omega\right) ^2}{2} }\cdot\dfrac{\left( -\; _{\alpha}\hat{d}\; t^{\alpha}\; \Omega\right)^m}{\sqrt{m!}}
\end{equation}
which is formally equivalent to a Poisson probability amplitude. \\

\noindent The use of the properties of the $_{\alpha}\hat{d}$  operator finally yields  the probability distribution

\begin{equation}\begin{split}\label{prP}
& _{\alpha}p_m (t) =\mid \langle m\mid \Psi\;\rangle\mid^2=e^{-  \left( \; _{\alpha}\hat{d}\; t^{\alpha}\; \Omega\right) ^2}\cdot\dfrac{\left(  _{\alpha}\hat{d}\; t^{\alpha}\; \Omega\right)^{2m}}{m!}=
\dfrac{X^m}{m!}e_{m}^{(\alpha,\;2)}(-X), \\
& X=\left( t^{\alpha}\; \Omega\right) ^2,\\
& e_{m}^{(\alpha,\;2)}(-X) =\sum_{r=0}^{\infty}\dfrac{(-1)^{r}}{r!}\dfrac{\Gamma(2(r+m)+1)}{\Gamma(2(r+m)\alpha +1)}X^r, \quad \forall X,m,\alpha\in \mathbb{R}, \alpha\leq1
\end{split}\end{equation}
which evidently reduces to a Poisson amplitude for $\alpha =1$  and in which we have used  the following definition for the function $e_{m}^{(\alpha,\;2)}(-X)$

\begin{equation}\label{my}
e_{s}^{(\alpha,\;\beta)}(\xi) :=\sum_{r=0}^{\infty}\dfrac{\xi^{\;r}}{r!}\dfrac{\Gamma(\beta(r+s)+1)}{\Gamma(\beta(r+s)\alpha +1)},\;\;\; \forall \alpha,\beta\in \mathbb{R},\; \beta>0, \alpha>\dfrac{1}{\beta}
\end{equation}

It can be checked that the probability \eqref{prP} is properly normalized, being

\begin{equation}
\sum_{m=0}^{\infty} {} _{\alpha}p_m =1
\end{equation}

\begin{figure}
	\centering
	\includegraphics[width=0.5\linewidth]{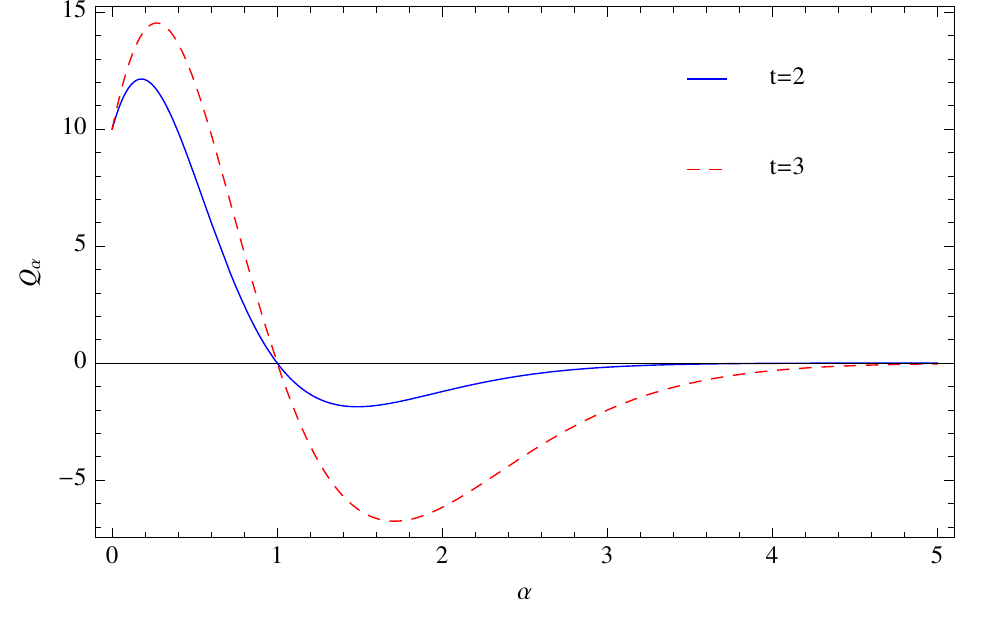}
	\caption{Mandel Parameter $Q_{\alpha}$ vs $\alpha$, for different values of $t$.}
	\label{Figure3}
\end{figure}

\begin{figure}[htp]
	\centering
	\begin{subfigure}[c]{0.48\textwidth}
		\includegraphics[width=0.9\linewidth]{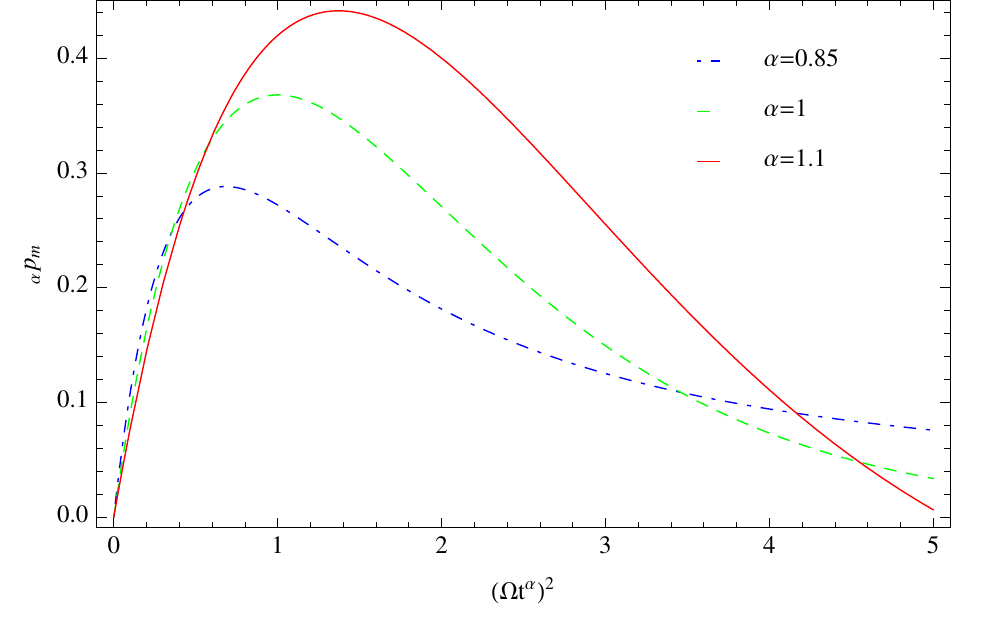}
		\caption{$m=1$.}
		\label{Fig3a}
	\end{subfigure}
	\begin{subfigure}[c]{0.48\textwidth}
		\includegraphics[width=0.9\linewidth]{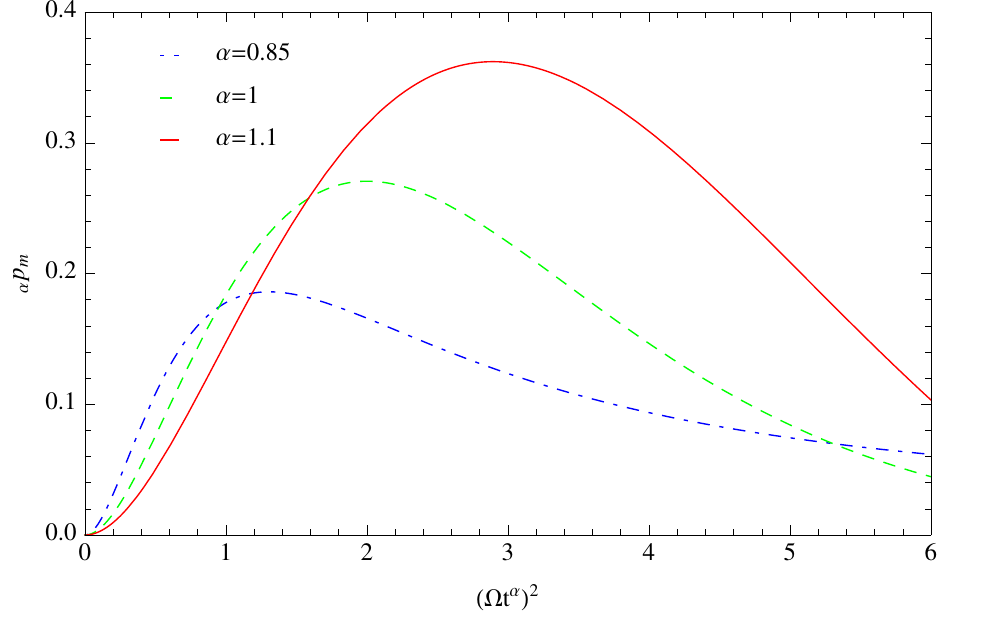}
		\caption{$m=2$.}
		\label{Fig3b}
	\end{subfigure}
	\begin{subfigure}[c]{0.48\textwidth}
		\includegraphics[width=0.9\linewidth]{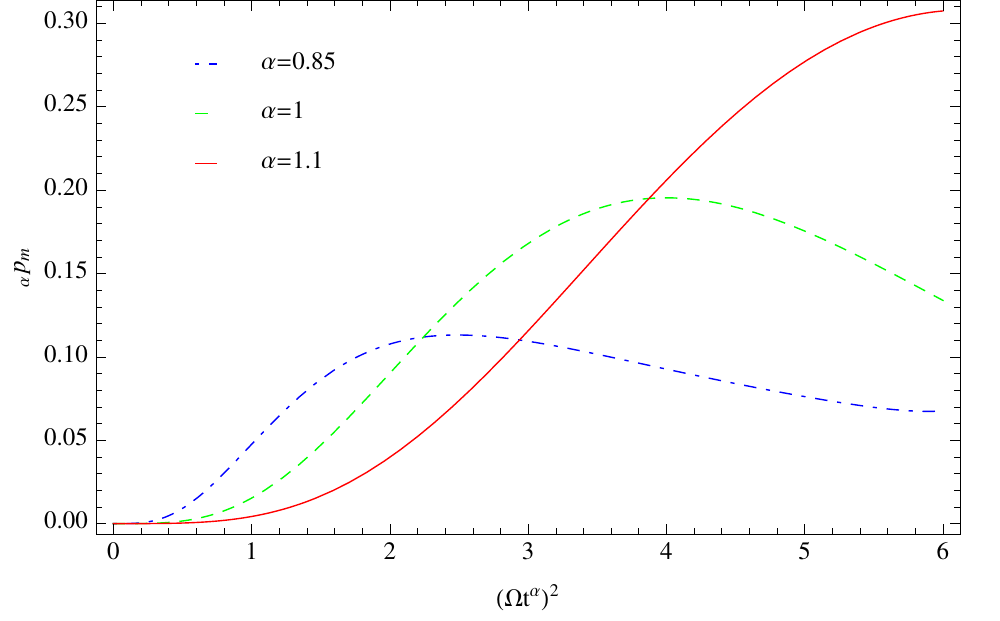}
		\caption{$m=4$.}
		\label{Fig3c}
	\end{subfigure}
	\\[3mm]
	\caption{Probability distribution "$  {}_{\alpha}p_{m}$" vs $\left( \Omega t^{\alpha}\right) ^2$.}\label{Figure4} 
\end{figure}

Furthermore regarding the evaluation of the average number of emitted photons we proceed as it follows

\begin{equation}\begin{split}\label{medPh}
 \langle \;m\;\rangle &=  \sum_{m=0}^{\infty} m\; {} _{\alpha} p_m=\sum_{m=0}^{\infty} m \dfrac{X^m}{m!}e_{m}^{(\alpha,\;2)}(-X)= \\
 & =e^{-\left( {}_{\alpha}\hat{d}^{\;2} X\right) }\sum_{m=1}^{\infty}\dfrac{\left( {}_{\alpha}\hat{d}^{\;2} X\right)^m}{(m-1)!}= {}_{\alpha}\hat{d}^{\;2} X=\dfrac{2X}{\Gamma(2\alpha+1)}
\end{split}\end{equation}
and an analogous procedure allows the evaluation of the r.m.s. of the emitted photons, namely

\begin{equation}\begin{split}\label{rms}
\sigma_{m}^{\;2}&=\langle\;m^2\;\rangle-\langle\;m\;\rangle^2\;={}_{\alpha}\hat{d}^{\;4} X^2+ {}_{\alpha}\hat{d}^{\;2} X-\left(\dfrac{2X}{\Gamma(2\alpha+1)} \right) ^2=\\
& = 2X\left[ 2X\left( \dfrac{6}{\Gamma(4\alpha+1)}-\dfrac{1}{\left( \Gamma(2\alpha+1)\right)^2 }\right)+\dfrac{1}{ \Gamma(2\alpha+1)}\right]  
\end{split}\end{equation}
We define the Mandel parameter

\begin{equation}
Q_{\alpha}=\dfrac{\sigma_{m}^{\;2}-\langle\;m\;\rangle}{\langle\;m\;\rangle}=2X\left( \dfrac{6}{\Gamma(4\alpha+1)}-\dfrac{1}{\left( \Gamma(2\alpha+1)\right)^2 }\right)\Gamma(2\alpha+1)
\end{equation}
The behaviour of $Q_{\alpha}$ vs. $\alpha$, for different values of $t$, is shown in Fig. \ref{Figure3}. Albeit this type of problems deserves a dedicated analysis, we point out that a process of photon emission ruled by a fractional Schr\"{o}dinger equation is fixed by a power law (recall that $X=\left( t^{\alpha}\Omega\right)^2 $ ), furthermore the presence of a region with $Q<0$
indicates the possibility of photon bunching. These are clearly pure speculations since the physical process ruled by  eq. \eqref{CrAnn} has not been defined.
The photon emission probability vs. $X$ and different $\alpha$  is given in Figs. \ref{Figure4}.\\

Further comments on the present results and the link with previous investigations will be provided in the following concluding section.

\section{Final Comments}

The process we have described so far and the derivation of the associated fractional Poisson probability amplitude are different from what is usually quoted in the literature.  Without entering into the phenomenology of the fractional Poisson processes we note that the equation governing the generating function of the distribution itself is given by \cite{Laskin}

\begin{equation}
G_{\alpha}(s,t)=E_{\alpha,1}\left(-(1-s)\;\Omega \;t^{\alpha} \right) 
\end{equation}
By the use of the umbral notation we can expand the previous generating function as

\begin{equation}
G_{\alpha}(s,t)=e^{\;_{\alpha}\hat{d}\;s\;(\Omega\; t^{\alpha})}e^{-\;_{\alpha}\hat{d}\;(\Omega\; t^{\alpha})}=\sum_{m=0}^{\infty}s^{m}\dfrac{_{\alpha}\hat{d}^{\;m}}{m!}(\Omega t^{\alpha})^{m}
\sum_{n=0}^{\infty}\dfrac{_{\alpha}\hat{d}^{\;n}}{n!}(-\Omega t^{\alpha})^{n}=\sum_{m=0}^{\infty}s^{m}{}_{\alpha}P_{m}(t)
\end{equation}
where

\begin{equation}\label{PoiPr}
{}_{\alpha}P_{m}(t)=\dfrac{(\Omega t^{\alpha})^{m}}{m!}\sum_{n=0}^{\infty}\dfrac{(n+m)!}{\Gamma(\alpha(n+m)+1)}\dfrac{(-\Omega t^{\alpha})^{n}}{n!}
\end{equation}
is the fractional Poisson distribution, introduced in refs. \cite{Zolotarev}, and derived here within the framework of our umbral formalism. According to the methods we have envisaged to calculate average and r.m.s. values, by setting
	
%

\begin{equation}
{}_{\alpha}P_{m}(t)= \dfrac{\left( _{\alpha}\hat{d}\;\Omega\;t^{\alpha}\right)^m }{m!}e^{-_{\alpha}\hat{d}\;\Omega\;t^{\alpha}}
\end{equation}
we find, for the first order moment,

\begin{equation}
\langle\;m\;\rangle= \sum_{m=0}^{\infty}m \dfrac{\left( _{\alpha}\hat{d}\;\Omega\;t^{\alpha}\right)^m }{m!}e^{-_{\alpha}\hat{d}\;\Omega\;t^{\alpha}} =
 \dfrac{\left(\Omega\;t^{\alpha} \right) }{\Gamma(\alpha+1)}  
\end{equation}
and, for the variance,

\begin{equation}\begin{split}
\sigma^{2}(t)=& \sum_{m=0}^{\infty}m^2 {}_{\alpha}P_{m}(t)-\left( \sum_{m=0}^{\infty}m\; {}_{\alpha}P_{m}(t)\right)^2 = \\
& = \sum_{m=1}^{\infty}m \dfrac{\left( _{\alpha}\hat{d}\;\Omega\;t^{\alpha}\right)^m }{(m-1)!}e^{-_{\alpha}\hat{d}\;\Omega\;t^{\alpha}}-\left( \dfrac{\left(\Omega\;t^{\alpha} \right) }{\Gamma(\alpha+1)}  \right)^2= \\
& =\dfrac{2 \left(\Omega\;t^{\alpha} \right)^2}{\Gamma(2 \alpha+1)}+\dfrac{\left(\Omega\;t^{\alpha} \right)}{\Gamma(\alpha+1)}-\dfrac{\left(\Omega\;t^{\alpha} \right)^2}{\left(\Gamma(\alpha+1) \right)^2 }
\end{split}\end{equation} 
in agreement with the results obtained in refs. \cite{Laskin,Zolotarev}.\\

We like to stress that this paper has clarified some questions regarding the handling of fractional $PDE$ of $M\!-\!L$ type and of the associated operator ordering. Most of the conclusions drawn here have required a critical understanding of the concept of semi-group properties associated with the $M\!-\!L$ function and has opened the way to some speculation yielding a different definition of the fractional Poisson distribution. \\

Before closing the paper a few left open points should be clarified.\\
We go back to the time fractional  Schroedinger reported in eq. \eqref{CrAnn} and write it using an hermitian version of the $\hat{H}$-operator, namely

\begin{equation}\begin{split}\label{newH}
& \left( i \partial_{t}\right)^{\alpha}\mid \Psi\;\rangle = \hat{H}\mid\Psi\rangle+i^{\alpha}\dfrac{t^{-\alpha}}{\Gamma(1-\alpha)}\mid 0\;\rangle, \\
& \hat{H}=\Omega\left( \hat{a}+\hat{a}^+\right) 
\end{split}\end{equation}
 The use of the ordering procedure outlined in the previous sections provides the solution of eq. \eqref{newH} in the form

\begin{equation}\label{newHsol}
\mid \Psi\;\rangle = e^{\;{}_{\alpha}\hat{d}\Omega(-it)^{\alpha}\left( \hat{a}+\hat{a}^+\right) }\mid 0\;\rangle= e^{\frac{(-1)^{\alpha}\left( {}_{\alpha}\hat{d}\;\Omega\; t^{\alpha}\right)^2 }{2}}\sum_{n=0}^{\infty}\dfrac{\left( (-i)^{\alpha}\Omega\; t^{\alpha}{}_{\alpha}\hat{d}\right)^n }{\sqrt{n!}}\mid n \;\rangle 
\end{equation}
The associated probability amplitude of finding $\mid \Psi \;\rangle $  into number state state reads

\begin{equation}
\langle m\mid \Psi \;\rangle = e^{\frac{(-1)^{\alpha}\left( {}_{\alpha}\hat{d}\;\Omega\; t^{\alpha}\right)^2 }{2}}\dfrac{\left( (-i)^{\alpha}\Omega\; t^{\alpha}{}_{\alpha}\hat{d}\right)^m }{\sqrt{m!}}
\end{equation}
thus finding for the for the square amplitude

\begin{equation}
\mid \langle m\mid \Psi\; \rangle\mid^2 = e^{\left( {}_{\alpha}\hat{d}\;\Omega\; t^{\alpha}\right)^2 \cos(\pi\;\alpha)} \dfrac{\left( \Omega\; t^{\alpha}{}_{\alpha}\hat{d}\right)^{2m} }{m!}
\end{equation}
which does not appear to be properly normalized.\\

It is finally worth mentioning that the coherent states, we have introduced in this paper, are different from those discussed in the second of ref. \cite{Laskin}, where they have been assumed to be provided by 

\begin{equation}
\mid \zeta \;\rangle = \sum_{n=0}^{\infty}\dfrac{\zeta^n}{\sqrt{n!}}e_n ^{(\alpha,1)}\left( -\dfrac{1}{2}\mid \zeta\mid^2\right)\mid n \;\rangle 
\end{equation}
where $\zeta$ is a complex quantity.\\
In the case discussed here we have assumed that they are generated via a Schroedinger like process of fractional type. \\

In conclusion, the article  has shown that a wise combination of umbral and operational methods may become quite a powerful tool to deal with problems emerging in fractional calculus, with the associated special functions and with the physical processes they may describe.  One of the main output of the  method, we have proposed, is the introduction in the procedure  of analytical means associated with the operator ordering, according to the suggestions put forward in the past  in ref. \cite{BD}.\\

In a forthcoming publications we will extend our point of view to problems requiring time ordering procedures \cite{Dattoli,Oteo} which will be handled by an appropriate redefinition of the Dyson series \cite{Dyson}.\\

\textbf{Acknowledgments} \\

The Authors express their sincere appreciation to Prof. \textbf{\textit{Vittorio Romano}}  for the kind hospitality at the University of Catania where this work was started and for suggesting ref. \cite{Uchaikin}.\\

K.G and A.H. were supported by the \textit{\textbf{PAN-CNRS}} program for French-Polish collaboration. Moreover, K.G. thanks for support from \textit{\textbf{MNiSW}} (Warsaw, Poland), "Iuventus Plus 2015-2016", program no IP2014 013073.\\

\textbf{References}\\

%
%
%
%
%
%

\end{document}